\documentclass[footinbib,aps,amsmath,amssymb,twocolumn,floatfix,prx,superscriptaddress,longbibliography]{revtex4-1}

\input{preamble.sty}

\AtBeginDocument{%
	\newwrite\bibnotes
	\def\bibnotesext{Notes.bib} 	\immediate\openout\bibnotes=\jobname\bibnotesext
	\immediate\write\bibnotes{@CONTROL{REVTEX41Control}}	\immediate\write\bibnotes{@CONTROL{%
			apsrev41Control,author="08",editor="1",pages="0",title="0",year="1"}}
 	\if@filesw
 	\immediate\write\@auxout{\string\citation{apsrev41Control}}%
	\fi
 }%
 
\begin{document}
	\title{Efficient multi-qubit subspace rotations via topological quantum walks}
	
	\author{Xiu Gu}
	\email{guxiu1@gmail.com}

	\affiliation{Tencent Quantum Laboratory, Tencent, Shenzhen, Guangdong 518057, China}
	
    \author{Jonathan Allcock}
    	\affiliation{Tencent Quantum Laboratory, Tencent, Shenzhen, Guangdong 518057, China}
    \author{Shuoming An}
    
    \affiliation{Tencent Quantum Laboratory, Tencent, Shenzhen, Guangdong 518057, China}
	  \author{Yu-xi Liu}
	\affiliation{School of integrated circuits, Tsinghua University, Beijing 100084, China}
    \affiliation{Frontier Science Center for Quantum Information, Beijing, China}
	\date{\today}
	
	\begin{abstract}
		The rotation of subspaces by a chosen angle is a fundamental quantum computing operation, with applications in error correction and quantum algorithms such as the Quantum Approximate Optimization Algorithm, the Variational Quantum Eigensolver and the quantum singular value transformation. Such rotations are usually implemented at the hardware level via multiple-controlled-phase gates, which lead to large circuit depth when decomposed into one- and two-qubit gates.  Here, we propose a fast, high-fidelity way to implement such operations via topological quantum walks, where a sequence of single-qubit $z$ rotations of an ancilla qubit are interleaved with the evolution of a system Hamiltonian in which a matrix $A$ is embedded. The subspace spanned by the left or right singular vectors of $A$ with non-zero singular values is rotated, depending on the state of the ancilla. This procedure can be implemented in superconducting qubits, ion-traps and Rydberg atoms with star-type connectivity, significantly reducing the total gate time required compared to previous proposals.

	\end{abstract}

\maketitle




\section{Introduction}


Multi-qubit subspace rotations are ubiquitous in quantum algorithm and ciruit design. Consider the multiple-controlled phase gate ${\rm{C}}_{n}\rm{Z}$. This is the gate acting on $n+1$ qubits which effects the reflection $$\mathbb{I}-2|1^{n+1}\rangle\langle 1^{n+1}|,$$ with $|1^{n+1}\rangle:=\ket{1}^{\otimes n +1}$. ${\rm{C}}_{n}\rm{Z}$ is locally equivalent to the generalized Toffoli gate (i.e., multiple-controlled NOT gate), and is a key component in Grover search~\cite{Nielsen2000}, Hamiltonian simulation~\cite{Poulin2018}, error correction~\cite{Paetznick2013,Schindler2011a,Dennis2001,Cory1998,Shor1995a,Bonesteel2012} and quantum factoring~\cite{Nielsen2000,Koley2012}. Viewing this reflection as a rotation of $\ket{1^{n+1}}$ by angle $\pi$, ${\rm{C}}_{n}\rm{Z}$ is a special case of the general subspace rotation $$\Pi_\phi=\exp(i2\phi\Pi),$$ 
which imparts a phase factor $e^{i2\phi}$ to the subspace defined by projector $\Pi$.  Such subspace rotations have widespread application, including in implementations of the Quantum Approximate Optimization Algorithm (QAOA) \cite{Farhi2014,Hill2021} 
, the Variational Quantum Eigensolver (VQE) \cite{Peruzzo2014,OMalley2016a,Nam2020,Gard2020}, and the Quantum Singular Value Transformation (QSVT) ~\cite{Gilyen2019,Martyn2021}. 

The efficient hardware implementation of subspace rotations is thus an important task, particularly in the Noisy Intermediate-Scale Quantum (NISQ) era of quantum computing \cite{Alexeev2021}.  However, conventional methods for decomposing ${\rm{C}}_{n}\rm{Z}$ gates and subspace rotations into single- and two-qubit gates can lead to long circuit depths~\cite{Barenco1995}, which can limit the kinds of algorithms that can be executed.  Previous proposals for more efficient implementations of ${\rm{C}}_{n}\rm{Z}$ have been based on tailored hardware design \cite{Khazali2020,Rasmussen2020,Banchi2015,Zhu2003,Wang2001,Katz2022,Fedorov2011, Nikolaeva2021, Xing2021,Kim2021,Hill2021,Chu2021,Banchi2015} and rely on mechanisms specific to certain physical platforms, or via complicated optimal control pulses~\cite{Zahedinejad2015,Zahedinejad2016}.

In this paper, we propose an efficient way to implement ${\rm{C}}_{n}\rm{Z}$ gates, as well as multi-qubit subspace reflections and rotations. Our approach is based on alternating a many-body-interaction with single-qubit rotations, and can be viewed as gate compilation via quantum walk. The required many-body interactions are native to a number of physical processors, and 
can be implemented in superconducting qubits~\cite{Kjaergaard2020,Blais2021,Wendin2016,Gu2017,Krantz2018}, ion-traps~\cite{Monroe2021,Bruzewicz2019} and Rydberg atoms~\cite{Saffman2010,Morgado2021,Eldredge2017,Zeytinoglu2022} with star-type connectivity, significantly reducing the total gate time required compared to previous proposals. For processors where each qubit has four nearest neighbours (such as in proposals for scalable superconducting architectures), our approach can be used to implement $\rm{C_3Z}$ as well as the four qubit $\Pi_\phi$ rotation about the subspace with projector $\Pi=\ketbra{0000}{0000}$.

At a high level, our procedure works as follows. The native many-body interactions used in our procedure have the effect of embedding a matrix $A$ in a block Hamiltonian $H$. Interleaving $e^{-iHt}$ and single-qubit $z$ rotations of an auxiliary qubit then digitally simulates the evolution of a discrete-time quantum walk, where the walk sequence
corresponds to a single-particle topological band model traversing the Brillouin zone~\cite{Kitagawa2010,Ramasesh2017,Flurin2016}. When such a quantum walk refocuses to its initial position after adiabatically traversing the Brillouin zone~\cite{Ramasesh2017,Flurin2016}, it acquires a Berry phase of either $0$ or $\pi$ depending on the corresponding topological winding number. In the case of the digitally simulated walk, with the auxiliary qubit initialized in  the $|0\rangle$ ($|1\rangle$) state, the right (left) singular vectors of $A$ with non-zero singular values acquire a Berry phase of $\pi$, whereas singular vectors of $A$ with zero singular value acquire a phase determined entirely by the sequence of single-qubit $z$ rotations. By choosing an appropriate set of single-qubit rotations, we are able to effect a desired relative phase shift between the subspaces corresponding to the zero and non-zero singular values of $A$.

The article is organized as follows. In \secref{sec:topology}, we review the discrete-time quantum walk and its 
connections to the Su-Schrieffer-Heeger (SSH) topological band model~\cite{Kitagawa2010,Asboth2016}. In particular, we highlight the existence of two distinct topological phases determined by the angle of the quantum walk \emph{coin}. The topological winding numbers of the two phases are imprinted on the Berry phase acquired by the quantum walk wavefunction, which we will use as a means to implement subspace rotations. In \secref{sec:stepDependent} we discuss how to digitally simulate such topological quantum walks via a simple interleaved sequence of unitary operators. In \secref{sec:matrixEmbedding} we show how to decompose a Hilbert space into a direct sum of subspaces corresponding to the topological walk phases via a matrix embedding. In \secref{sec:cQED} and \secref{sec:trappedIon}, we show how to implement subspace rotations in superconducting circuits, benchmark our approach with a previous proposal~\cite{Fedorov2011}, and investigate its robustness to anharmonicity. In \secref{sec:trappedIon} we discuss the implementation of our method in trapped-ion and Rydberg-atom systems. We end with a discussion and conclusions in~\secref{sec:discussion} and \secref{sec:conclusion}.

\section{Topology of quantum walks}
\label{sec:topology}

Let us first review the discrete-time quantum walk and its connection to topological insulators.  This has been the subject of a number of previous works~\cite{Kitagawa2010,Ramasesh2017,Flurin2016}, and here we summarize the key concepts.

In a discrete-time quantum walk, one considers a principal system -- the \emph{walker} -- supplemented with an additional two-level \emph{coin} system.  At each time step, the coin state is rotated, and the position of the walker translated in a direction determined by the coin.  This single-step evolution can be expressed as the unitary
\be 
W_{0}(\theta)=S_0R(\theta), \label{eq:singleStep}
\ee
where $R(\theta)$ is a parameterized rotation of the coin and
\be
S_0=\exp(i\hat{k}\sigma_{z})\label{eq:momentum}.
\ee
Here, $\sigma_{z}$ is the Pauli $z$ operator acting on the coin, and $\hat{k}$
is the momentum operator of the walker, with eigenvalues $k$ in the continuous range $[0,2\pi]$. In this paper we will take 
\bea
R(\theta)&=\cos(\theta/2)I-i\sin(\theta/2)\sigma_x,\label{eq:coinToss} \nonumber \\
&= \bpm \cos(\theta/2) & -i\sin(\theta/2) \\-i\sin(\theta/2) & \cos(\theta/2)\epm
\eea 
where $\sigma_{x}$ is the Pauli $x$ operator of the coin, i.e. a rotation about the $x$ axis, where $\theta$ determines the probability of obtaining heads ($|1\rangle$) or tails ($|0\rangle$).

\eqref{eq:singleStep} can be expressed as the evolution of an effective Hamiltonian over time $\delta t$, 
\be
W_{0}(\theta)=\exp [-iH_{\rm eff}(k,\theta)\delta t],
\ee 
where
\be
H_{\rm eff}(k,\theta)=\int_{-\pi}^{\pi}dk(E_{k,\theta}\vec n_{k,\theta} \cdot\vec{\sigma})\otimes|k\rangle\langle k|,\label{eq:Ek}
\ee
$\vec{\sigma} = (\sigma_x,\sigma_y,\sigma_z)$ is the vector of Pauli matrices acting on the coin space, and $ E_{k,\theta}$ and $\vec n_{k,\theta}$ are defined by
\bea 
&&\cos E_{k,\theta}=\cos k\cos\frac{\theta}{2},\label{eq:nk}\\
&&\vec n_{k,\theta}=\lp\cos k\sin\frac{\theta}{2},-\sin k\sin\frac{\theta}{2},\sin k\cos\frac{\theta}{2}\rp/\sin E_{k,\theta}.\nonumber
\eea
Note that above and throughout this paper we work in units where $\delta t=1$ and set $\hbar=1$. 


From \eqref{eq:Ek}, one can see that the Hamiltonian decomposes into a direct sum of two-by-two blocks labeled by $k$. 
In other words, $H_{\rm eff}$ depicts a two-band model, where the eigenstate at momentum $k$ of the upper (lower) band is the $\vec{n}_{k,\theta}$-up ($\vec{n}_{k,\theta}$-down) state.

\begin{figure}[hbt]
	\includegraphics[width=8.5cm]{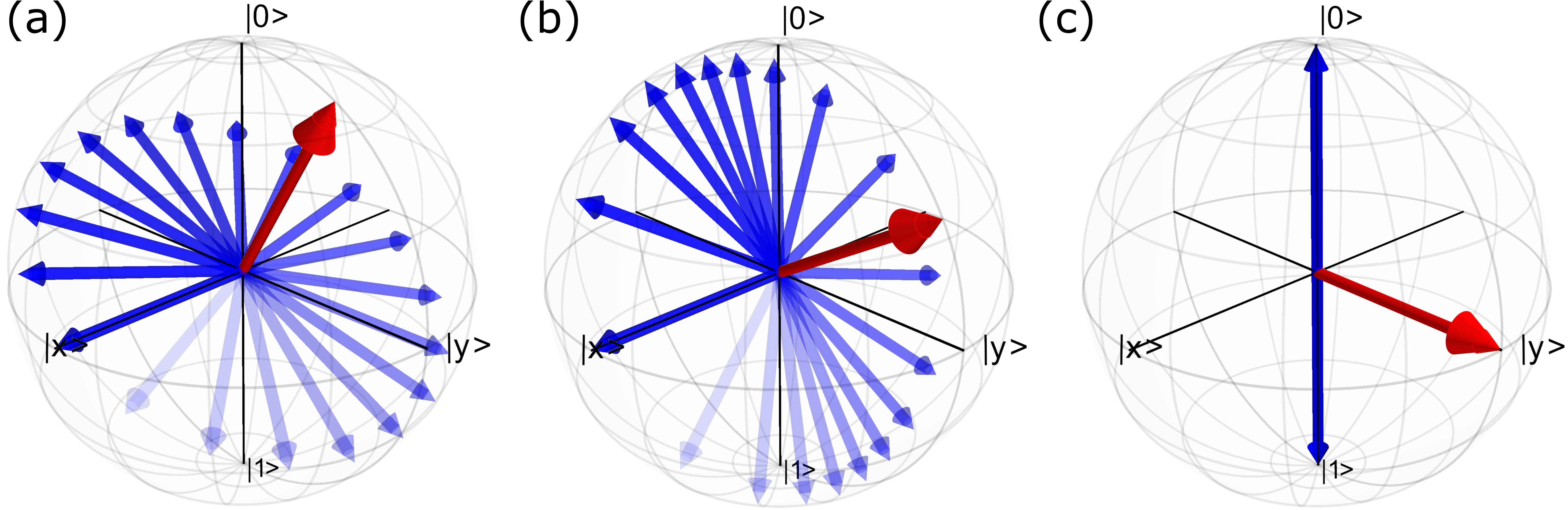}
	\caption{ Schematic evolution of $\vec{n}_{k,\theta}$ (blue vector) as $k$ moves from $0$ to $2\pi$ (where the blue color changes from dark to light).  (a, b) Topological phase with winding number 1. $\vec{n}_{k,\theta}$ traces out a great circle perpendicular to $\vec{A}=(0,\cos\frac{\theta}{2},\sin\frac{\theta}{2})$ (red vector),  as $k$ runs from $[0,2\pi]$. (a) $\theta=2/3 \pi$. (b) $\theta=1/3 \pi$. (c) The topologically trivial phase corresponds to $\theta=0$, for which $\vec{n}_{k,\theta}$ coincides with z axis and the winding number is 0. 
		\label{fig:Bloch}}
\end{figure}

$H_{\rm eff}$ is equivalent to the Su-Schrieffer-Heeger (SSH) topological-band model~\cite{Asboth2016} which is characterized by an integer-valued topological invariant, either 0 or 1, corresponding to the number of times $\vec n_{k,\theta}$ winds around the origin as $k$ is varied adiabatically from $0$ to $2\pi$.
$H_{\rm eff}$ has a unitary chiral symmetry of the form 
\be
e^{i\pi\vec{A}_\theta \cdot \vec{\sigma} } H_{\rm eff} (k,\theta) e^{-i\pi \vec{A}_\theta \cdot \vec{\sigma} }=-H_{\rm eff}(k,\theta),
\label{eq:chiral}
\ee
where $\vec{A}_\theta=(0,\cos\frac{\theta}{2},\sin\frac{\theta}{2})$ is perpendicular to $\vec n_{k,\theta}$ for all $k$~\cite{Kitagawa2010}. Since chiral symmetry constrains $\vec n_{k,\theta}$ to lie on a great circle perpendicular to $\vec{A}_\theta$, there are two distinct topological phases depending on the value of $\theta$:
\begin{enumerate}[i]
    \item For $\theta\neq 0,2\pi$, the winding number is 1, corresponding to the topological phase, as shown in Figs. \ref{fig:Bloch} (a) and (b). 
    
    \item For $\theta= 0,2\pi$, $\vec n_{k,\theta}$ coincides with the $z$ axis, corresponding to the trivial phase, with winding number 0, as shown in \figref{fig:Bloch} (c). 
\end{enumerate}

\section{Step-dependent topological quantum walks}
\label{sec:stepDependent}

The topological phases are nonlocal and are not easy to probe. However, the topological winding number is imprinted on the Berry phase acquired by the quantum walk wavefunction as the walker traverses adibatically across the Brillouin zone. 
This process of $k$ sweeping from $0$ to $2\pi$ can be digitially simulated by choosing an integer $N$ and defining the $m$th-step quantum walk unitary as \cite{Cedzich2013,Ramasesh2017,Flurin2016}
\be
W_{\delta k}(m)=S^mW_{0}(\theta),
\ee where $S=\exp(i\delta \hat{k}\sigma_z)$, $ \delta =2\pi/N$ and $W_{0}(\theta)$ is as defined in \eqref{eq:singleStep}. In terms of the effective Hamiltonian of \eqref{eq:Ek}, this is
\be
W_{\delta k}(m)=\exp [-iH_{\rm eff}(k+m\delta k,\theta)], \label{eq:Trotter}\ee 
i.e, in the $m$th step of the modified walk, $k$ changes to $k+m\delta k$.
Thus, simulating the evolution of $k$ from $0$ to $2\pi$ is achieved by a sequence of steps from $1$ to $N$, each step interleaving $W_0$ and $S^m$:
\be
W_{ k}^{[N,1]}(\theta):=W_{\delta k}(N)W_{\delta k}(N-1)\ldots W_{\delta k}(1), \label{eq:WNQ}
\ee
By \eqref{eq:Trotter}, this is a Trotterized version of the evolution of a time-dependent Hamiltonian~\cite{Ramasesh2017,Flurin2016}

\be W_{ k}^{[N,1]}(\theta)\approx \exp{\{-i\int_{0}^{N}H_{\rm{eff}}(k+\delta k(t),\theta)dt\}}\label{eq:Walkdynamics},
 \ee
where $\delta k(t)=\sum_{m=1}^N \Theta(t-m)2\pi/N$, with $\Theta$  the step function.  See detailed analysis in~\cite{Ramasesh2017,Flurin2016}. 

As the system evolves under  \eqref{eq:Walkdynamics}, 
the axis $\vec{n}_{k,\theta}$ (\eqref{eq:nk}) changes adiabatically to $\vec{n}_{k+\delta k(t),\theta}$, and traces out a closed path when $\delta k(t)$ completes a cycle from $0$ to $2\pi$.
The eigenstates of the walker and coin transform as

\bea
\ket{k}\otimes |n_{k,\theta}{\rm -up}({\rm down})\rangle&\ra \nonumber\\
\ket{k+2\pi}\otimes  |n_{k+2\pi,\theta}{-\rm up}({\rm down})\rangle,
\eea
i.e. return to their original states, but acquire an overall phase with dynamical and geometric contributions. This revival of Bloch oscillations due to topological winding has been demonstrated experimentally in cold atoms~\cite{Atala2013} and superconducting circuits~\cite{Flurin2016}. As the $\vec{n}_{k,\theta}$-up and $\vec{n}_{k,\theta}$-down states have opposite energies $E_{k,\theta}$ (\eqref{eq:nk}), they accumulate dynamical phases with opposite signs. On the other hand, the geometric phase is connected to the Berry phase acquired by the two-level system during the adiabatic cyclic evolution~\cite{Sakurai1994,Leek2007}, and is proportional to the solid angle of the cone subtended by the closed path traced out by the quantization axis $\vec{n}_{k,\theta}$ when $\delta k(t)$ completes a cycle from $0$ to $2\pi$. When $\theta=0, 2\pi$ (trivial phase), $\vec{n}_{k,\theta}$ coincides with the $z$ axis, and the geometric phase acquired is zero.  When $\theta\neq 0,2\pi$ (topological phase), as mentioned in the previous section, the chiral symmetry (\eqref{eq:chiral}) constrains $\vec{n}_{k,\theta}$ to lie on a great circle perpendicular to $\vec{A}$. Therefore, all $k$ states acquire the same geometric phase $\pi$.



The competing influences of the dynamical and geometric phase on the quantum walk wavefunction causes \eqref{eq:WNQ} to differ in an important way depending on the value of $\theta$:

\begin{enumerate}[i)]
\item $\theta=0,2\pi$, i.e.~the topologically trivial phase. In this case, the coin rotation operator reduces to the identity operator $R(\theta)=\mathbb{I}$, and \eqref{eq:WNQ} simplifies to a product of $z$ rotations, giving
\eql{
W_{ k}^{[N,1]}&=S^{(N+1)N/2}S_0^{N},&\quad N \text{ even} \nonumber\\
W_{ k}^{[2N,1]}&=S_0^{2N}, &\quad N \text{ odd} 
\label{eq:trivial}
}
where we used the fact that $S^N=\mathbb{I}$ but, in general, $S^{(N+1)N/2}\neq\mathbb{I}$. 

\item  $\theta\neq 0,2\pi$, i.e.~the topological phase. In this case, the behaviour of the walk is captured quantitatively by the revival theorem~\cite{Cedzich2013}:
\eql{
||W_{ k}^{[N,1]}+(-1)^{N/2}\mathbb{I}||&=2|\cos({\theta}/{2})|^{N/2}, &\quad N \text{ even}\nonumber\\
||W_{ k}^{[2N,1]}+\mathbb{I}||&=2|\cos({\theta}/{2})|^{N}. &\quad N\text{ odd} \label{eq:revival}
}
\end{enumerate}

By taking $N$ to be odd and large enough that $\abs{\cos(\theta/2)}^N \ll 1$, one can thus engineer 
\eql{
W_{ k}^{[2N,1]}(\theta)&=S_0^{2N},&\quad \theta=0,2\pi \label{eq:summary1}\\ 
W_{ k}^{[2N,1]}(\theta)&\approx -\mathbb{I},&\quad \theta\neq 0, 2\pi 
\label{eq:summary2}
}
causing the relative evolution of the trivial and topological phases to be $-S_0^{2N}$, fully controlled by $S_0$ (\eqref{eq:momentum}).

Equations~(\ref{eq:summary1}) and (\ref{eq:summary2}) provide a route to implementing subspace rotations. Decomposing a Hilbert space of interest into a direct sum of subspaces corresponding to quantum walks with different coin parameters $\theta$, the operator sequence \eqref{eq:WNQ} will cause the $\theta\neq 0$ subspaces to acquire a geometric phase $\pi$, while the $\theta=0$ subspaces will be rotated by $S_0^{2N}$.

In the next section we shall show how such a direct sum decomposition is naturally achieved via matrix embedding.

\section{Subspace rotation via matrix embedding}

\label{sec:matrixEmbedding}
Following Lloyd et al.~\cite{Lloyd2021}, we consider a Hamiltonian which embeds a matrix $A$ as
\be
H=\sigma^+A+\sigma^-A^\dagger,\label{eq:JC}
\ee 
where $\sigma^\pm$ are the raising and lowering operators for an auxiliary qubit.

Writing the singular value decomposition of $A$ as
\be
A=\sum_{J}\Lambda_{J}\left|l_{J}\right\rangle \left\langle r_{J}\right|,\ee
where $\Lambda_{J}$ are the singular values of $A$, and $|l_{J}\rangle $ ($|r_{J}\rangle )$ are the corresponding left (right) singular vectors, the time evolution of the system takes the form
\bea
&&\exp{(-iHt)}=\exp[-it(\sigma^{+}A+\sigma^{-}A^{\dagger})]\label{eq:HmatrixElement}\\
&&=\bigoplus_{J}\left[\begin{array}{cc}
	\cos(\Lambda_{J}t)\left|l_{J}\right\rangle \left\langle l_{J}\right| & -i\sin(\Lambda_{J}t)\left|l_{J}\right\rangle \left\langle r_{J}\right|\\
	-i\sin(\Lambda_{J}t)\left|r_{J}\right\rangle \left\langle l_{J}\right| & \cos(\Lambda_{J}t)\left|r_{J}\right\rangle \left\langle r_{J}\right|
\end{array}\right].\nonumber
\eea
where the matrices are written in the $\{\ket{1},\ket{0}\}$ basis of the auxiliary qubit, ordered such that the top left matrix element corresponds to $\ketbra{1}{1}$ (i.e. we are expressing $e^{-iHt}$ in the $\{\ket{1}\ket{l_J},\ket{0}\ket{r_J}\}$ basis). The Hilbert space thus decomposes into a direct sum of $2\times2$ blocks labeled by the singular values of $A$. Comparison with \eqref{eq:coinToss} shows that each block can be viewed as implementing a separate quantum walk coin with rotation angle $\theta/2 = \Lambda_jt$. 

As a concrete example of such a matrix embedding, one can think of \eqref{eq:JC} as the famous Jaynes-Cummings (JC) model \cite{Haroche2006}, where a two-level atom interacts with a quantized light field. In that case, the embedded matrix $A$ corresponds to the photon annihilation operator $a=\sum_n \sqrt{n}|n-1\rangle\langle n|$,  with $n$ the photon number. The $2\times2$ blocks (\eqref{eq:HmatrixElement}) are defined through the conservation of total excitations, and spanned by the basis {$|1\rangle|n-1\rangle$,$|0\rangle|n\rangle$}. The singular values $\Lambda_{J}$ correspond to the square root of the photon number. Note that $a$ has a $0$ singular value with corresponding left (right) singular vector $|0\rangle$ ($|N\rangle$), where $N$ is the dimension of the truncated space.

By the arguments in \secref{sec:topology} and \secref{sec:stepDependent}, if one defines the single-step unitary 
\bea
W_0(k,t):&=&S_0\exp{(-iHt)}\nonumber\\
&=&\exp{(i\sigma_z k)}\exp[-it(\sigma^{+}A+\sigma^{-}A^{\dagger})]\\
&=&R_z(2k)\exp[-it(\sigma^{+}A+\sigma^{-}A^{\dagger})],\label{eq:W0}
\eea
where $R_z(\theta)=e^{i\frac{\theta}{2}\sigma_z}$ is a single qubit $Z$ rotation, then the sequence 
\eql{
W_k^{[2N,1]}(t)&=S^{2N}W_0(k,t)\ldots S^2W_0(k,t)SW_0(k,t), \label{eq:sequence}
}
where $N$ is odd and $S=\exp{(i\sigma_z2\pi/N)} = R_z(4\pi/N)$ is the $z$ rotation of the ancilla,  implements a direct sum of topological quantum walks of ~\eqref{eq:WNQ}, where the behaviour of each block differs depending on the value of $\Lambda_J t$:
\begin{widetext}
\eql{
W^{[2N,1]}_k(t)&\approx
\bigoplus_{J:\Lambda_Jt\neq 0,2\pi}\left[\begin{array}{cc}
	-\left|l_{J}\right\rangle \left\langle l_{J}\right| & 0\\
	0& -\left|r_{J}\right\rangle \left\langle r_{J}\right|
\end{array}\right] 
&+\bigoplus_{J:\Lambda_Jt = 0,2\pi}\left[\begin{array}{cc}
	e^{2iNk}\left|l_{J}\right\rangle \left\langle l_{J}\right| & 0\\
	0& e^{-2iNk}\left|r_{J}\right\rangle \left\langle r_{J}\right| 
\end{array}\right] \label{eq:approxR}
}
\end{widetext}

By \eqref{eq:trivial} and \eqref{eq:revival}, the second term on the right hand side of \eqref{eq:approxR} above is exact, while the approximation in the first term depends on the value of $2\left|\cos(\Lambda_{J}t)\right|^N$.  For $N$ large enough that the approximation above holds to within a desired tolerance, when the ancilla is initialized in the $|1\rangle$ ($|0\rangle$) state, the subspace spanned by the left (right) singular vectors $|l_{J}\rangle$ ($|r_{J}\rangle$) of $A$ acquire a phase factor of $-1$ when $\Lambda_jt\neq 0, 2\pi$, and a phase factor of $e^{2iNk}$($e^{-2iNk}$) when $\Lambda_Jt =0,2\pi$.

Singular vectors of $A$ with zero singular value therefore always acquire a phase of $e^{\pm 2iNk}$. For singular vectors with $\Lambda_J\neq 0$, if one can find a time $t$ such that $\abs{\cos(\Lambda_{J}t)}$ is simultaneously small for all $J$, then only a small number of steps $N$ is required for all of these singular vectors to aquire a phase factor of $-1$, and one can efficiently implement a subspace rotation between the subspaces spanned by the zero and non-zero singular vectors of $A$.  

In the next section we will show how this approach can be implemented in state-of-the-art quantum processors.  Before doing so, let us make a number of remarks.

First, compared with \eqref{eq:momentum} where $S_0=e^{i\hat{k}\sigma_z}$ is expressed in terms of the momentum operator $\hat{k}$, in \eqref{eq:W0} it is sufficient for our purposes to consider the case where  $S_0=e^{ik\sigma_z}$, i.e. where the walker initially populates a single $k$ component.  At each step, the walker's position is shifted by $\pm k$ depending on the coin state.

Second, in the case where $A$ is Hermitian, \eqref{eq:JC} reduces to $\sigma_x\otimes A$ and, by performing a basis change on the ancilla, i.e. $\sigma_x\rightarrow\sigma_z$, $\sigma_y\rightarrow- \sigma_y$, $\sigma_z\rightarrow\sigma_x$, one can implement subspace rotations of the form~\eqref{eq:approxR} via a quantum walk sequence (c.f. ~\eqref{eq:sequence}) consisting of alternating single qubit $\sigma_x$ rotations and $\sigma_z\otimes A$ interactions. This is similar to the geometric phase gates proposals~\cite{Wang2001,Wang2002,Zhu2003,Pechal2012,Cross2015} where, instead of employing a digital walk sequence, the transformations were achieved through continuously changing couplings between the system qubits and an auxiliary harmonic mode. 

\section{circuit QED implementation}
\label{sec:cQED}
Superconducting circuits~\cite{Kjaergaard2020,Blais2021,Wendin2016,Gu2017,Krantz2018,You2011} are among the leading platforms for quantum computers, and well-suited to implementing our proposal.  We show how this can be done without additional hardware requirements in a superconducting system where a central ancilla qubit is coupled to four neighbors. Reflection and rotation operations are implemented by a topological quantum walk sequence, which alternately applies simultaneous ${\rm{C}}\rm{Z}$ gates \cite{Gu2021} between the ancilla and neighbor qubits, and single-qubit $z$ rotations of the ancilla.  

\begin{figure}[b]
	\includegraphics[width=8.3cm]{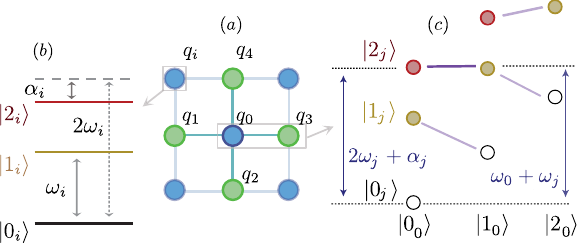}
	\caption{(a) Superconducting qubits (circles) with nearest-neighbor coupling (lines) form a grid lattice.  (b) The energy levels of of qubit $i$. $|0_i\rangle$ is the ground state  and $|1_i\rangle$ and $|2_i\rangle$ are the first and second excited states with frequency $\omega_i$, and $2\omega_i+\alpha_i$ , where $\alpha_i $ is the anharmonicity. (c) CZ gates between qubit $0$  and qubit $j$ through the resonant Rabi oscillation between $|2_j0_0\rangle$ and $ |1_j1_0\rangle\ $ (dark purple lines).  Other transitions (light purple lines) are far off-resonance. 
		\label{fig:energylevel} }
\end{figure}
 We consider a square-lattice set-up as shown in \figref{fig:energylevel} (a). 
The superconducting-qubit system is composed of a central qubit $q_0$ and $N_q=4$ nearest neighbor qubits. Each qubit $i$ has three energy levels $|0_i\rangle$, $|1_i\rangle$, $|2_i\rangle$, with frequency spacing $\omega_i$, and $2\omega_i+\alpha_i$, where $\alpha_i $ is the anharmonicity (\figref{fig:energylevel} (b)).  
The system Hamiltonian in the laboratory frame reads
\bea
& &{H}_{\rm qlab}= H_0+H_{\rm int}\label{eq:lattice},\\
& &H_0=\sum_{i=0}^{N_q}\left(\omega_{i} b_{i}^{\dagger} b_{i}+\frac{\alpha_{i}}{2} b_{i}^{\dagger} b_{i}^{\dagger} b_{i} b_{i}\right)\label{eq:H0},\\ 
& & H_{\rm int}=\sum_{i=1}^{N_q}\left(g_{i0}b_{i}b_{0}^{\dagger}+g_{i0}b_{0}b_{i}^{\dagger}\right)\label{eq:Hint} ,
\eea
where $b_i^\dagger$ ($b_i$) is the bosonic creation (annihilation) operator for the $i$th superconducting qubit, $\omega_i$ is the frequency, $\alpha_i$ is the detuning, and $g_{i0}$ is the coupling strength between central qubit $q_0$ and its neighbor $q_i$. In our analysis, we ignore the coupling between qubits positioned on the diagonals of the square lattice, i.e. $q_i$ and $q_j$, with $i,j\neq0$.

\subsection{Many-body interactions through simultaneous CZ gates}
As depicted in \figref{fig:energylevel} (c), a CZ gate can be applied between ancilla $q_0$ and any neighbor $q_j$,  by tuning  $|0_02_j\rangle$ into resonance with $ |1_01_j\rangle\ $ \cite{Strauch2003,Barends2014,Negrneac2020,Sung2020}, i.e. by taking
\be
\omega_{0}=\omega_{j}+\alpha_{j}.\label{eq:SimuCZ}\ee
If we set the Rabi frequency of resonant transition $|0_02_j\rangle\leftrightarrow |1_01_j\rangle\ $ to $1$, state $ |1_01_j\rangle\ $ acquires a minus sign after gate time $\pi$.
This CZ interaction can be activated by directly tuning the qubit frequency~\cite{DiCarlo2009,Negrneac2020} or can be turned on and off through an additional coupler~\cite{Yan2018,McKay2016,Arute2019,Sung2020}.

In the following, we consider running these resonant CZ gates between $q_0$ and all its neighboring qubits simultaneously~\cite{Gu2021}. For simplicity, we ignore the finite time required to turn these interactions on or off, and model these interactions of simultaneous CZ gates as ideal rectangular pulses.

Switching to the interaction picture with respect to $H_0$ (\eqref{eq:H0}), the interaction Hamiltonian $H_{\rm int}$ (\eqref{eq:Hint}) under the simultaneous CZ gates operation (\eqref{eq:SimuCZ})  can be simplified to,
 \be
 H_q=\sigma_{0}^{10}\sum_{i=1}^{N_q}g_{i}\sigma_{i}^{12}+\sigma_{0}^{01}\sum_{i=1}^{N_q}g_{i}\sigma_{i}^{21},\label{eq:sys}
 \ee
where $\sigma_{k}^{ij}=|i\rangle\langle j|_{k}$, $g_i=\sqrt{2} g_{i0}$ is the coupling strength between resonant energy levels
$|0_{0}2_{i}\rangle$ and $|1_{0}1_{i}\rangle$. The factor $\sqrt{2}$ comes from the fact that transmon qubits are very close to being harmonic, and the interaction involves two excitations. We use the rotating wave approximation (RWA) in obtaining \eqref{eq:sys}, and
neglect exchange interaction terms $\sigma_{0}^{10}\sigma_{i}^{01}+\rm{H.c.}$ and 
 $\sigma_{0}^{21}\sigma_{i}^{01}+\rm{H.c.}$ (the light purple transitions in \figref{fig:energylevel} (c)), which are detuned by $|\alpha_i|$ and $|\alpha_i+\alpha_0|$. This approximation is valid in the large dispersive regime corresponding to heavy detuning  $|\alpha_i| \gg g_{i0}$.
We shall later go beyond the RWA and analyze the full Hamiltonian~\eqref{eq:lattice}, accounting for finite anharmonicity. 

From a physics perspective, \eqref{eq:sys} can be understood as a variant of the Dicke model~\cite{Klimov2009,Tsyplyatyev2010}, where 
a single harmonic photon mode (corresponding to qubit $q_0$) is coupled to many two-level atoms (corresponding to the neighbour qubits, with the ground and excited states of the atoms mapped to qubit states $|1\rangle$, $|2\rangle$). In our case, due to the heavy detuning, at most a single excitation (photon) of $q_0$ is allowed, and conservation of total excitation number in the Dicke model constrains the whole dynamics to the single-excitation and  zero-excitation subspaces. 

In \eqref{eq:sys}, as the photon mode $q_0$ only couples to the $|1\rangle$, $|2\rangle$ states of its neighbors, 
the number of neighbours initialized in state $\ket{0}$ affects the number ($D$) of atoms which $q_0$ couples to. 
To see this, consider different initial states for $N_q=4$ neighbour atoms. Under the condition required for our simultaneous CZ gates (\eqref{eq:SimuCZ}):
\begin{enumerate}[i)]
\item An initial state $|1_01100\rangle$ is on-resonance with $|0_02100\rangle$ and $|0_01200\rangle$, while the next-nearest state  $|0_01110\rangle$ ($|0_01101\rangle$) is detuned with $|1_01100\rangle$ by $\alpha_3$ ($\alpha_4$). Under the RWA, only the resonant interaction is considered (\eqref{eq:sys}). So when the system is initialized in state $|1_01100\rangle$, it experiences Rabi oscillations described by the Hamiltonian  
\be
{\Lambda_{1100}}|1_0\rangle\langle 0_0|\otimes |l_{1100}\rangle\langle  r_{1100}|+\rm{H.c} \label{eq:1100},
\ee
where $|l_{1100}\rangle=|1100\rangle$, $\langle r_{1100}|=({g_1}\langle 2100|+{g_2}\langle 1200|)/{{\Lambda_{1100}}}$, and $\Lambda_{1100}=\sqrt{g_1^2+g_2^2}$ is the effective Rabi frequency. This corresponds to the single excitation subspace of the inhomogeneous Dicke model where central mode $q_0$ couples to $D=2$ atoms, $q_1$ and $q_2$, with coupling strengths $g_1$ and $g_2$.

\item When the initial state is $|1_01110\rangle$, this corresponds to the Dicke model with $D=3$ atoms. Projected on to $|1_01110\rangle$,  \eqref{eq:sys} takes the form
\be
{\Lambda_{1110}}|1_0\rangle\langle 0_0| \otimes|l_{1110}\rangle\langle  r_{1110}|+\rm{H.c}  \label{eq:1110}
\ee
where  $|l_{1110}\rangle=|1110\rangle$, 
$\langle r_{1110}|=({g_1}\langle 2110|+{g_2}\langle 1210|+{g_3}\langle 1120|)/{{\Lambda_{1110}}}$, $\Lambda_{1110}=\sqrt{g_1^2+g_2^2+g_3^2}$ is the effective Rabi strength.
\item Similar statements can be made, for instance, when the initial state is $\ket{1_0 0000}$ ($D=0$), $\ket{1_0 1000}$ ($D=1$), or $\ket{{1_01111}}$ ($D=4)$.
\end{enumerate}
In summary, \eqref{eq:sys} describes the Dicke model with $D$ atoms, where $D$ is the number of neighbour qubits initialized in state $\ket{1}$. We list different representative states for $D=0,1,2,3,4$ in the second column of \tabref{tab:states}. Since we are only interested in computational states, initial states with $|2\rangle$ populated are excluded. We show the resonant (and next nearest) states in column three (four) when the simultaneous CZ gates operation is activated. The detuned states are ignored in the RWA. We will discuss  off-resonace couplings to these detuned states in~\secref{sec:anharm}. 
\begin{table}[h]
\begin{tabular}{>{\centering}p{0.2cm}>{\centering}p{2.2cm}>{\raggedright}p{3cm}>{\raggedright}p{3cm}}
	\toprule 
      D  & representative states & resonant states& states detuned by $\alpha$\tabularnewline
	\midrule
	\midrule 
	0  & $|0_{0}0000\rangle$ & none & none\tabularnewline
	\midrule 
	0& $|1_{0}0000\rangle$ & none & $|0_{0}1000\rangle$, $|0_{0}0100\rangle,$ $|0_{0}0010\rangle$,
	$|0_{0}0001\rangle$\tabularnewline
	\midrule 
	1 & $|1_{0}1000\rangle$ & $|0_{0}2000\rangle$ & $|0_{0}1100\rangle$, $|0_{0}1010\rangle$, $|0_{0}1001\rangle$ \tabularnewline
	\midrule 
	2 & $|1_{0}1100\rangle$ & $|0_{0}2100\rangle$, $|0_{0}1200\rangle$ & $|0_{0}1110\rangle$, $|0_{0}1101\rangle$ \tabularnewline
	\midrule 
	3 & $|1_{0}1110\rangle$ & $|0_{0}2110\rangle$, $|0_{0}1210\rangle$, $|0_{0}1120\rangle$ & $|0_{0}1111\rangle$ \tabularnewline
	\midrule 
	4 & $|1_{0}1111\rangle$ & $|0_{0}2111\rangle$, $|0_{0}1211\rangle$, $|0_{0}1121\rangle$,
	$|0_{0}1112\rangle$ & none\tabularnewline
	\bottomrule
\end{tabular}
	\caption{Representative states for the effective Dicke model with $D$ atoms when the simultaneous CZ gates operation of \eqref{eq:sys} is activated. As we are only interested in computational states, the initial states with $|2\rangle$ populated are excluded. The third and fourth column shows the
		resonant and next nearest states with the representative states in the first column. For simplicity we assume all the qubits have the same anharmonicity $\alpha$ in the fourth column.
		\label{tab:states}}
\end{table}

With an understanding of the dynamics of the simultaneous CZ gates, let us now reformulate \eqref{eq:sys} in the context of \secref{sec:matrixEmbedding} by introducing an embedded matrix $A$:

\be
A=\sum_i^{N_q}g_i\sigma_i^{12},\label{eq:embedA}\ee 
which has singular value decomposition
\be
A = \sum_{J\in\{0,1\}^{N_q}} \Lambda_J\ketbra{l_J}{r_J},
\ee
where the indices $J\in\{0,1\}^q$ are bit strings of length $N_q$.

Let us first take a look at the nonzero singular values of $A$, which is equivalent to $D\neq0$ in the Dicke model.
The corresponding left and right singular vectors are given by
\bea
\ket{l_J}= \ket{J} \label{eq:lnon0}\\
\ket{r_J}= \sum_{i: J_i=1} g_i/\Lambda_J\ket{J_{\uparrow i}}
\eea
where $J$ is a bit string that has at least one bit of $1$, $J_i$ is the $i$th bit of $J$, and $J_{\uparrow_i}$ denotes the binary string equal to $J$, except on the $i$th coordinate where it is equal to $2$. 
So the right singular vectors $|r_J\rangle$ are out of the computational space, and $|r_J\rangle$ should be empty before and after the gate sequence.
The corresponding singular value $\Lambda_{J}$ satisfies
\be
\Lambda_J^2={\sum_{i=1 }^{N_q}J_ig_{i}^{2}},\label{eq:RabiFreq}
\ee
This is the same with the single excitation subspaces of the Dicke model, with $D=\sum_iJ_i$. For example, taking $N_q=4$ neighbors and $J=1110$ (\eqref{eq:1110}), the corresponding singular value $\Lambda_J=\sqrt{g_1^2+g_2^2+g_3^2}$ is the Rabi oscillation frequency between states 
$|1_0\rangle|l_{1110}\rangle$, 
and $|0_0\rangle| r_{1110}\rangle$. 

Let us analyze $D=0$, which maps to the zero singular value $\Lambda_{\bar{0}}=0$ of $A$, where $\bar{0} = 0^{N_q}$, the all zero state.
It has the corresponding left singular vector 
\be
|l_{\bar{0}}\rangle=|0_1 0_2 0_3...0_{N_q}\rangle \label{eq:l0}\ee
The corresponding right singular vector $|r_{\bar{0}}\rangle$ is not written out explicitly since it is not in the computational space.

Taking the ancilla $q_0$ into account, for example, see \eqref{eq:1100} and \eqref{eq:1110}, the action of the simultaneous CZ gates Hamiltonian $H_q$ can be expressed as
\bea
H_q|1_0,l_{J}\rangle&=&\Lambda_{J}|0_0,r_{J}\rangle,\nonumber\\
H_q|0_0,r_{J}\rangle&=&\Lambda_{J}|1_0,l_{J}\rangle,
\eea 
and thus acts as a $\sigma_x$ operation in the
$\{|1_0,l_{J}\rangle,|0_0,r_{J}\rangle\}$ basis. The evolution of the different spaces labelled by $J$ are Rabi oscillations
\be
\exp{(-iH_qt)}=\bigoplus_{J}\left[\cos(\Lambda_{J}t)-i\sin(\Lambda_{J}t)\sigma_{x}\right],\label{eq:Rabi}
\ee where $\sigma_x$ is defined in the $|1_0,l_{J}\rangle$, $|0_0,r_{J}\rangle$ basis. Note that for the zero singular value $\Lambda_{\bar{0}}=0$, $|1_0,l_{\bar{0}}\rangle$ is the dark state, which does not evolve under $H_q$.


\subsection{Subspace rotations}
The computational space of the neighbor qubits are spanned by the left singular vectors $|l_J\rangle$ given in~\eqref{eq:lnon0} and~\eqref{eq:l0}, i.e. where $J$ are bit strings with either at least one $1$, or all bits $0$. When the native many-body interactions are activated by the simultaneous CZ gates, these computational states $|l_J\rangle$ undergo different dynamics. As shown in \eqref{eq:Rabi}, when the ancilla is initialized in $|1\rangle$, the left singular vector $|l_{\bar{0}}\rangle$,  i.e. with all the neighbour qubits in the ground states, corresponding to $0$ singular value, remains unchanged; while left singular vectors $|l_{J\neq\bar{0}}\rangle$,
i.e. with at least one of the neighbour qubits flipped from 0 to 1, oscillate at frequency $\Lambda_J$.


The dynamics of the simultaneous CZ gates can be viewed in the context of topological walks. Comparison to~\eqref{eq:HmatrixElement} shows that the rotation of different subspaces labelled by $J$ in~\eqref{eq:Rabi} can be viewed as coin space rotations, and thus we can interleave the native many-body interaction with single qubit $z$ rotations of the ancilla to implement subpace rotations.
\begin{figure*}[hbt]
	\includegraphics[width=16cm]{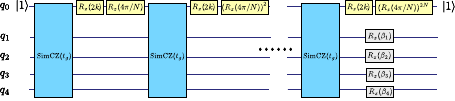}
	\caption{Gate sequence to implement subspace rotation $\Pi_\phi=\exp(i2\phi\Pi)$, where $\Pi=|0000\rangle\langle 0000|$ acts on qubits $q_1,q_2,q_3,q_4$. The ancilla $q_0$ is initialized in state $|1\rangle$ and, after the topological quantum walk sequence, returns to state $|1\rangle$. ${\rm SimCZ}(t_g)$ is the simultaneous CZ gates operation between $q_0$ and $q_i$, by setting $\omega_0=\omega_i+\alpha_i$
		$i=1,2,3,4$.  Here $\omega_i$ ($\alpha_i$) is the $i$th qubit frequency (anharmonicity). The gate time $t_g$ is chosen to minimize the number of steps $N$. $R_z(\theta)$ denotes a single qubit $z$ rotation. The angle $k$ determines the subspace rotation angle $\phi$. Detailed analysis see \eqref{eq:sequenceCircuit}. The final set of single qubit $z$ rotations paramaterized by $\beta_i$ (grey boxes) are chosen to account for the change from the interaction picture to the laboratory frame.\label{fig:gateseq} }
\end{figure*}
The sequence of gates used to implement subspace rotations in superconducting circuits is shown in~\figref{fig:gateseq} and, from~\eqref{eq:W0}, corresponds to
\bea
& &W_{k}^{[2N,1]}(t_g)=S^{2N}W_0(k,t_g)\ldots SW_0(k,t_g), \label{eq:sequenceCircuit}
\eea
where $W_0(k,t_g)=R_0^z(2k)\exp{(-iH_qt_g)}$, $S=R_0^z(4\pi/N)$, and
 \be
R_0^z(\theta)=e^{i\frac{\theta}{2}\sigma_0^{z}}
\ee
is the single qubit $z$ rotations of ancilla $q_0$. $H_q$ is defined by the simultaneous CZ gates as in \eqref{eq:sys}, which have a duration of $t_g$.

By~\eqref{eq:approxR}, if we initialize the ancilla $q_0$ to $|1\rangle$, it will return to state $|1\rangle$ after the topological sequence, $|l_{\bar{0}}\rangle$ will acquire a phase of $e^{ 2iNk}$, while $|l_{J\neq\bar{0}}\rangle$ will aquire a phase factor of $-1$.  

For $k=0$, this sequence of gates effects the reflection
\be 
2|l_{\bar{0}}\rangle \langle l_{\bar{0}}|-\mathbb{I}.
\ee
This gate operation inverts the phase of the target qubit conditioned on all the control qubits being $0$ (instead of $1$), which is equivalent to the ${\rm C}_3{\rm Z}$ gate up to a global phase $-1$, and local transformations.

For general $k$, this realizes a projector-controlled phase shift~\cite{Martyn2021} \be
\Pi_\phi=\exp(i2\phi\Pi),\ee 
where 
\be
\phi=\pi-Nk,
\ee
 and the projector $\Pi=|\bar{0}\rangle\langle\bar{0}|$. For $N_q=4$, this $4$-qubit $\Pi_\phi$ operator is locally equivalent to the triple-controlled phase gate $\rm{C}_3\rm{PHASE}(2\varphi)$, the four qubit operator which effects the transformation
\eql{
	\ket{0000}&\rightarrow e^{2i\phi}\ket{0000}\\
	\ket{ijkl}&\rightarrow\ket{ijkl}\qquad (i,j,k,l)\neq (0,0,0,0)
}
and which has applications in applying the Quantum Approximate Optimization Algorithm (QAOA)~\cite{Farhi2014} to Boolean satisfiability problems. For instance, in Ref.~\cite{Hill2021} the double-controlled phase $\rm{C}_2\rm{PHASE}$ gate was applied to the MAX-3-SAT problem, and $\rm{C}_3\rm{PHASE}$ (and, more generally $\rm{C}_{k-1}\rm{PHASE}$) can similarly be applied to MAX-4-SAT (MAX-k-SAT) problems. 

\subsection{Benchmarking}\label{sec:benchmark}
Here we analyze the performance of this approach in the ideal case, where the Hamiltonian is given by~\eqref{eq:sys}. In~\secref{sec:anharm} we will consider the full Hamiltonian~\eqref{eq:lattice} and model the effect of anharmonicity.


\subsubsection{Gate fidelity}
We first calculate the average gate fidelity $F$ of implementing subspace rotations by our approach, assuming ideal single-qubit gates. From Ref.~\cite{Nielsen2002}, this is given by 
\be
F=\frac{\abssq{\tr{M U_{\rm ideal}^\dag}} + \tr{M^\dag M}}{n(n+1)}, \label{eq:nielsen-fidel}\ee
where $n $ is the dimension of the computational space, $ U_{\rm ideal}=\Pi_\phi$ is the target gate and 
\be M=\langle1_0|W_{ k}^{[2N,1]}(t_g)|1_0\rangle,
\ee
since the ancilla is initialized in $|1\rangle$ and returns to $|1\rangle$ with probability 1 in the ideal case.


Here we consider the case $k=0$, i.e., a reflection. In \appref{app:rotation-fidelity} we consider more general $k$.  When $k=0$, $U_{\rm ideal} = \mathsf{diag}(1,-1,-1\ldots, -1)$, and $M$ can be expressed as $M = \bigoplus_J \bpm 2\cos(\Lambda_J t)^{2N}-1\epm$ (see~\appref{app:poly-transform}). It follows that
\begin{align}
\tr{M^\dag M} &= \sum_{J} \lp 2 \cos(\Lambda_J t_g)^{2N}-1\rp^2,\\
\tr{MU^\dag_{\rm ideal}}&=1 - \sum_{J\neq \bar{0}}\lp 2 \cos(\Lambda_J t_g)^{2N}-1\rp,
\end{align}
where the $\Lambda_J$ are given by~\eqref{eq:RabiFreq}. For concreteness, we take $q_0$ connected to 
$N_q=4$ neighbors (\figref{fig:energylevel} (a)) for which the dimension of the computational space is $n=16$. Results are described below for both homogeneous and inhomogeneous couplings, and summarized in Table~\ref{tab:ideal-results}.

\noindent\textbf{Homogeneous couplings.} We first consider the case where $g_i=g$, giving $\Lambda_J/g=\{1,\sqrt{2},\sqrt{3},2\}$. By \eqref{eq:revival}, the error is bounded by $2|\cos(\Lambda_J t_g)|^N$ and therefore the larger $N$ is, the better the achievable gate fidelity. The gate time $t_g$ is chosen such that $\abs{\cos(\Lambda_{J}t)}$ is simultaneously small for all $J$, 
which minimizes the $N$ needed to reach high fidelity.
We choose $t_g=0.333\pi/g$, such that $|\cos(\Lambda_J t_g)| \lesssim 0.5$. 
With $N=3$, the average gate fidelity is $F=0.980$, while $N=5$ and $N=7$ give $F=0.999$. 
\\
\noindent\textbf{Inhomogeneous couplings.} Next we consider inhomogeneous coupling strengths which may arise, for instance, from imperfections in fabrication and electronic control.  We set $g_i/g\approx\{0.85, 0.99, 0.91, 1.02\}$, which are generated by a Gaussian distribution $\mathcal{N} (1,0.1^2)$. By taking $t_g=0.333\pi/{\rm{max}} (g_i)$, the same calculations show we can reach an average gate fidelity $F=0.999$ with $N=7$. 
\begin{table}[h]
	\begin{tabular}{>{\centering}p{2cm}>{\centering}p{1.5cm}>{\centering}p{1.5cm}>{\centering}p{2cm}}
	\toprule 
     & N=3 & N=5 & N=7 \tabularnewline
     \midrule
     \midrule 
\text{homogeneous}    & 0.9804 & 0.9988 & 0.9999\tabularnewline
\midrule
\text{inhomogeneous} & 0.9721 & 0.9974 & 0.9997\tabularnewline
\bottomrule
    \end{tabular}\caption{Average gate fidelities for  subspace reflections ($k=0$), for $N_q=4$ neighbour qubits coupled to a central ancilla, assuming both homogeneous and inhomogeneous couplings.  }\label{tab:ideal-results}
\end{table}


\subsubsection{Gate time}


We now compare the time and resources required to implement our approach with other methods. The total gate time of both ${\rm{C}}_{3}\rm{Z}$ and $\Pi_\phi$ are approximately the same in our approach, since we can merge the single-qubit rotations $S_0$ and $S$. For homogeneous couplings, taking $N=5$ translates to 10 single-qubit rotations of $q_0$, and a total time performing simultaneous two-qubit gates of around $3.33 \pi$, with Rabi frequency $g$ of the $|02\rangle\leftrightarrow |11\rangle\ $ transition set to $1$. In contrast, the time to implement a single CZ gate is $\pi$. 

As a benchmark, we consider an efficient ${\rm{C}}_{2}\rm{Z}$ gate realization proposed by Fedorov, Steffen, Baur, da Silva and Wallraff (FSBSW)~\cite{Fedorov2011}, where states are transferred back and forth between non-computational and computational space. 
In \appref{app:cnz}, we show how the FSBSW method can be generalized to implement ${\rm{C}}_{n-1}\rm{Z}$ at a total CZ gate time cost of $(2n-3)\pi$. In addition, a single qubit rotation on each of the qubits is required at the end of the protocol to compensate for dynamical phases acquired (as is required, for instance, when tuning two-qubit CZ gates~\cite{Negrneac2020}).   


In contrast, assuming all-to-all coupling, decomposition of a $4$-qubit Toffoli gate (locally equivalent to ${\rm{C}}_{3}\rm{Z}$) into single- and two-qubit gates (`$1,2$Q') requires 13 two-qubit gates \cite{Barenco1995}. 

\begin{table}[h]
	\begin{tabular}{>{\centering}p{2cm}>{\centering}p{2cm}>{\centering}p{2cm}>{\centering}p{2cm}}
		\toprule 
		${\rm C_{3}Z}$ & This work & FSBSW
		~\cite{Fedorov2011} & $1,2$Q
		~\cite{Barenco1995}\tabularnewline
		\midrule
		\midrule 
		two-qubit gate time & $3.33$ CZ & 5 CZ & 13 two-qubit gates\tabularnewline
		\midrule 
		single-qubit gate time  & 10 $R_{z}$ & 1$R_{z}$ & not counted\tabularnewline
		\bottomrule
	\end{tabular}\caption{Time cost required to implement ${\rm C_{3}Z}$. }\label{tab:cost}
\end{table}

These costs are summarized in~\tabref{tab:cost}. In contrast to the FSBSW and $1,2$Q, our  approach requires the least two-qubit gate time, an advantage given that two-qubit gates are more costly to implement and more error-prone than single-qubit gates. Furthermore, we will show in~\secref{sec:discussion}
that by a slight modification to our protocol, we can implement a  ${\rm{C}}_{5}\rm{Z}$ gate with same time required for ${\rm{C}}_{3}\rm{Z}$ i.e. $3.33$ CZ. In comparison, the same gate via the FSBSW approach requires $9$ CZ gate time. However, the method of FSBSW has the advantage that it is implementable in systems of qubits arranged in a linear chain with nearest-neighbor coupling. Our protocol, on the other hand, is native to star-type connections, where several qubits share a common qubit.

\subsection{Effect of anharmonicity}\label{sec:anharm}
In $\secref{sec:cQED}$ we have so far performed calculations in the RWA, ignoring leakage into non-computational states. We now consider the general Hamiltonian $\eqref{eq:lattice}$, with each qubit's Hilbert space limited to three levels,  and investigate the effects of finite anharmonicity by taking the anharmonicities of $q_0, \ldots, q_4$ to be
\be
\alpha/(2\pi)=\{-262 , -249, -0.283, -295, -290\}{\rm MHz} ,\label{eq:alpha}
\ee which are generated by a Gaussian distribution around $-300$, with $10\%$ deviation. 
We set the frequency of the central ancilla qubit to be 
$\omega_0/(2\pi)=5.15$ GHz, with other qubit frequencies determined  by \eqref{eq:SimuCZ} during the simultaneous CZ gates operations. We assume homogenous couplings  $g_i=g$  between the center and neighbor qubits
(which can be achieved, e.g., by a tunable coupler~\cite{Yan2018}).  The simultaneous CZ gates are turned on and off as ideal rectangular pulses. 



\begin{figure}[h]
	\includegraphics[width=8cm]{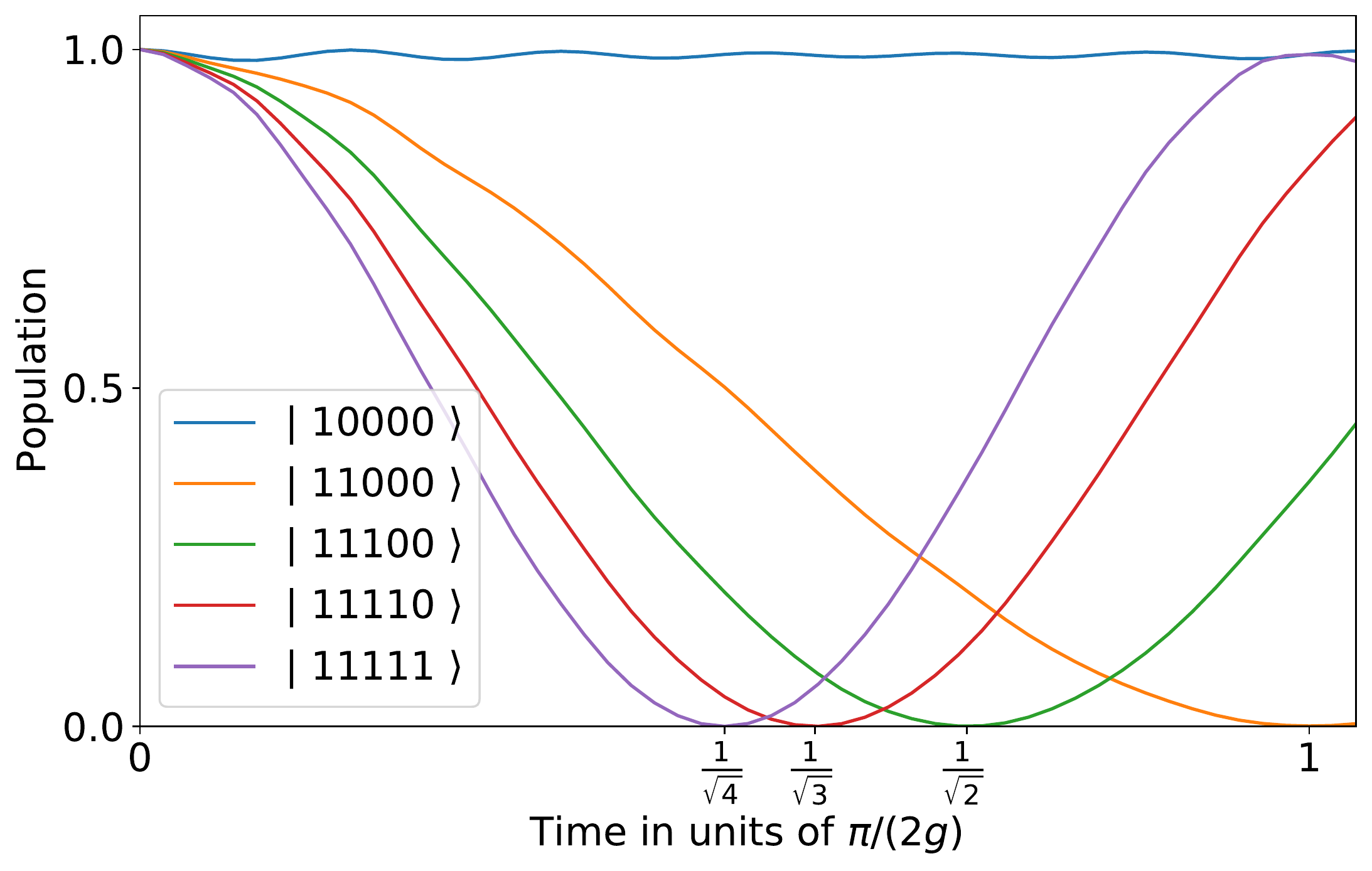}
	\caption{ Evolution of different initial states under simultaneous CZ gates, where a central ancilla $q_0$ is coupled to four physical qubits with coupling strength $g/2\pi=9$MHz. The anharmonicities are $\alpha/(2\pi)=\{-262 , -249, -0.283, -295, -290\}{\rm MHz}$ and $|\alpha/g|_{\rm min}=28$. Apart from the slight oscillations of the $\ket{10000}$ state, anharmonicity has little effect on the other states.
		\label{fig:energylevelprobe} }
\end{figure}

\noindent\textbf{Simultaneous CZ gates.} We first examine the evolution of different initial states under the simultaneous CZ gates. As shown in \figref{fig:energylevelprobe}, 
when the system is initialized in state $|10000\rangle$,  the population of $|10000\rangle$ (blue line) oscillates slightly around $1$ due to leakage to off-resonant states, see \tabref{tab:states}.
When the system is initialized in state $|11000\rangle$ (orange line), it corresponds to the Dicke model with $1$ atom. By \eqref{eq:RabiFreq},  at time $\pi/(2g)$ state $|11000\rangle$ is swapped to $|02000\rangle$, and the population of $|11000\rangle$ reduces to $0$.
Similarly, when the system is initialized in the state corresponding to the Dicke model with $D$ atoms (see \tabref{tab:states}), the state population is swapped to $0$ at time $\pi/(\sqrt{D}2g)$. 

Note that, while here we model each qubit's Hilbert space as being limited to three levels, in general multi-photon transitions to even higher levels may be possible if there is accidental resonance with these higher levels. However, we rule out any significant effect due to these multi-photon transitions by repeating the evolution for the specific input states of~\figref{fig:energylevelprobe} with an increased local Hilbert space dimension, and find that there is little qualitative difference even when increasing each qubit's Hilbert space dimension from three to nine.


\noindent\textbf{Subspace rotation/reflection.} Next we simulate the sequence of interleaving the simultaneous CZ gates with single qubit gates on $q_0$, as in 
\eqref{eq:sequenceCircuit}. For simplicity in 
$W_0$ we set $k=0$, giving $R_0^z(2k)=\mathbb{I}$. 
In the previous analysis, the simultaneous CZ gates operation in $W_0$ was calculated in the interaction picture using $H_q$ of \eqref{eq:sys}. Here we replace $H_q$ with the full Hamiltonian of \eqref{eq:lattice}, which takes into account all the off-resonance couplings and takes place in the lab frame. In this case, the free Hamiltonian $H_0$ in \eqref{eq:lattice} adds single-qubit phases to the ancilla $q_0$ and its neighbors. The extra phase on $q_0$ can be absorbed into the definition of $R_0^z(2k)$, and thus a non-zero net subspace rotation angle may result even with $k=0$.
To account for the extra phases on the neighbors, at the end of the sequence we need to apply single qubit rotations on the neighbor qubits,
\begin{equation}
	U_{\text {phase}} \equiv R^z_{1}\left(\beta_{1}\right) \otimes R^z_{2}\left(\beta_{2}\right) \otimes R^z_{3}\left(\beta_{3}\right)\otimes R^z_{4}\left(\beta_{4}\right),
\end{equation}
where $R^z_i$ denotes a single-qubit rotation about the z axis for qubit $i$, and the angles $\beta_i$ are chosen to correct for the change of frame.


In \secref{sec:benchmark}, the gate fidelity was calculated using a two-level approximation of ancilla $q_0$ assuming $|\alpha_i| /g=\infty$. In this section, all the qubits are modelled as three-level systems and, to go beyond the RWA, we vary $|\alpha_i| /g$ in the following way.  We consider $g/2\pi$ from the set $g/2\pi=\{9,3,2\}$MHz, corresponding to ratios of $\abs{\alpha/g}$ of
$$\abs{\alpha/g}_{\min}: = \min_{i}\abs{\alpha/g}\in \{28,83, 125\}$$
The results, calculated with Qutip~\cite{Qutip1}, are summarized in \tabref{tab:fidelity} for different walk steps $N$, and show that for large but finite values of anharmonicity, high fidelities can still be achieved. As is expected, the larger $|\alpha_i| /g$ the better the gate fidelity.  In particular, when $|\alpha_i| /g\approx 125$, we obtain a fidelity of $F=0.9945$ for $N=5$. 
 However, as the effect of leakage becomes more pronounced when $|\alpha_i| /g$ is small, longer quantum walk sequences do not guarantee better gate fidelity. Such leakage to non-computational space is also detrimental to the FSBSW and $1,2$Q approaches. While the DRAG~\cite{Motzoi2009} method can be used to combat leakage in single-qubit gates, it is unclear whether its efficacy generalizes to the multiqubit setting.
 
 

\begin{table}[h]
\begin{tabular}{>{\centering}p{1.5cm}>{\centering}p{1.5cm}>{\centering}p{1.2cm}>{\centering}p{1.2cm}>{\centering}p{1.2cm}>{\centering}p{1.2cm}}
	\toprule 
     	$g/2\pi$MHz& $|\alpha/g|_{\rm min}$   & N=3 & N=5 & N=7&$\phi$\tabularnewline
	\midrule
	\midrule 
 9& $\infty$ & 0.9804 & 0.9988 &0.9999& $\pi$\tabularnewline
		\midrule 
	2& 125 &  0.9780& \textbf{0.9945} & 0.9943& 3.061 \tabularnewline
		\midrule 
	3& 83 &  0.9758& \textbf{0.9888}& 0.9870&  3.021\tabularnewline
	\midrule 
	9& 28 & \textbf{0.9531} &  0.9348 & 0.8983& 2.893 \tabularnewline

	\bottomrule
\end{tabular}\caption{Average gate fidelity of quantum walk implementation of $\rm\Pi_{\rm{\phi}}$ in the presence of anharmonicity, as a function of the number of walk steps $N$ and $|\alpha/g|_{\rm min}$. The subspace rotation angle $\phi$, in the sixth column corresponds to the maximum fidelity case (bold) in each row. 
Anharmonicities for qubits $q_0,\ldots, q_4$ are $\alpha/(2\pi)=\{-262 , -249, -0.283, -295, -290\}{\rm MHz}$. The coupling strength $g_i$ between qubits $q_0$ and $q_i$ is set at $g_i=g$ in the first column. Off-resonant couplings are ignored in the first row,  which corresponds to the ideal RWA case with $k=0$, as calculated in \secref{sec:benchmark}. The rest are calculated with the full Hamiltonian in the lab frame. 
}\label{tab:fidelity}
\end{table}

\subsection{Discussions}

Instead of utilizing simultaneous CZ gates,  one might consider using simultaneous iSWAP gates \cite{Gu2021}. This has the effect of 
replacing the collective interaction in \eqref{eq:sys} by $\sum_i^{N_q}g_i\sigma_i^{01}$, i.e. transitions between computational states $|0\rangle$ and $|1\rangle$.
However, this goes beyond the single-excitation subspace of the Dicke model. In the four-atom homogeneous Dicke model, apart from state $|0000\rangle$, there are two additional dark states in the two-excitation subspace \cite{Klimov2009} and, as they belong to the zero-singular value subspace of $A$, will not be affected by the target reflection operation.  Thus the subspace rotated has dimension three, instead of one as in the $\rm{C}_n\rm{PHASE}$ gates. 

\section{Subspace rotations in trapped-ion and Rydberg systems}
\label{sec:trappedIon}
In this section we briefly discuss how our proposal for multi-qubit subspace rotations can be performed in trapped-ion~\cite{Monroe2021,Bruzewicz2019} and Rydberg systems~\cite{Saffman2010,Morgado2021,Eldredge2017,Zeytinoglu2022}. 

Trapped-ion systems~\cite{Monroe2021,Bruzewicz2019} are another popular candidate for quantum computers.
The workhorse in trapped-ion processors is the multi-ion entangling M{{\o}}lmer-S{{\o}}rensen (MS) gate~\cite{Molmer1999}, which can entangle up to 24 ions~\cite{Pogorelov2021}.
The unitary operation implemented by the MS gate is parametrized by two angles $\theta$ and $\varphi$,
\be
U_{\mathrm{MS}}(\theta,\varphi)=\exp\left(-\mathrm{i}\frac{\theta}{4}\left(\cos\varphi S_{x}+\sin\varphi S_{y}\right)^{2}\right),\ee
where ${S}_{x,y}=\sum_{i=0}^{N_q}\sigma_{i}^{x,y}$, with $\sigma_{i}^{x,y}$  the Pauli operators acting on the $i$th ion.
We denote ion number $0$ as the ancilla and define \cite{Muller2011}
\bea
    w_0&=&U_{\mathrm{MS}}(-\theta,0)\exp\left[\mathrm{i}\frac{\pi}{2}\sigma_{0}^{z}\right]U_{\mathrm{MS}}(\theta,0)\nonumber\\
 	&=&\exp\left[\mathrm{i}\frac{\theta}{4}\tilde{S}_{x}\sigma_{0}^{x}\right]\exp\left[\mathrm{i}\frac{\pi}{2}\sigma_{0}^{z}\right]\exp\left[-\mathrm{i}\frac{\theta}{4}\tilde{S}_{x}\sigma_{0}^{x}\right]\nonumber\\
	&=&\exp\left[\mathrm{i}\frac{\pi}{2}\left(\cos\left(\frac{\theta}{2}\tilde{S}_{x}\right)\sigma_{0}^{z}+\sin\left(\frac{\theta}{2}\tilde{S}_{x}\right)\sigma_{0}^{y}\right)\right]\nonumber\\
	&=&\bigoplus_{\lambda}|\lambda\rangle\langle\lambda|\otimes\nonumber\\
	& &\exp\left[\mathrm{i}\frac{\pi}{2}\left(\cos\left(\frac{\theta}{2}\lambda\right)\sigma_{0}^{z}+\sin\left(\frac{\theta}{2}\lambda\right)\sigma_{0}^{y}\right)\right] \label{eq:ion}
	\eea
where we use  $\left(S_x\right)^{2}=N_q\mathbb{I}+\sum_{ij}\sigma_{i}^{x}\sigma_{j}^{x}$ and ${\tilde S}_{x}=\sum_{i=1}^{N_q}\sigma_{i}^{x}$, i.e. excluding the ancilla. 

Here ${\tilde S}_{x}$ has eigenstates  ${\tilde S}_{x}|\lambda\rangle=\lambda|\lambda\rangle$.  For a given $|\lambda\rangle$, \eqref{eq:ion} is a rotation about axis $\vec{n}=(0,\sin(\theta\lambda/2),\cos(\theta\lambda/2))$  of ancila $q_0$. We can change \eqref{eq:ion} to an $x$ rotation via the identity  
\be
e^{\frac{i}{2}k\sigma_{x}}=-e^{\frac{i}{2}\pi(\cos\frac{k}{2}\sigma_{z}+\sin\frac{k}{2}\sigma_{y})}e^{\frac{i}{2}\pi\sigma_{z}}\label{eq:rotatex},
\ee
and define the walk unitary as 
\bea
W_0&=&S_0\left(w_0e^{i\pi/2\sigma_0^z}\right)\nonumber\\
&=&-\bigoplus_{\lambda}|\lambda\rangle\langle\lambda|\otimes S_0
 \left[\begin{array}{cc}
	\cos(\frac{\theta\lambda}{2}) & i\sin(\frac{\theta\lambda}{2})\\
i\sin(\frac{\theta\lambda}{2}) &\cos(\frac{\theta\lambda}{2})
\end{array}\right] \label{eq:ion-walk}
\eea
where $S_0=\exp{(i\phi\sigma_0^z)}$. Therefore, we can interleave $W_0$ and $S=\exp{(i2\pi/N\sigma_0^z)}$ to rotate subspaces labeled by zero eigenvalue of ${\tilde S}_{x}$, c.f. \eqref{eq:approxR}.  Note that since the left and right singular vectors coalesce into the same eigenvectors $|\lambda\rangle$, the initial state of the ancilla $q_0$ can be initialized in either state $|1\rangle$ or $|0\rangle$.  


Let us take a closer look at the eigenstates of ${\tilde S}_{x}$.  ${\tilde S}_{x}$ is equivalent to  ${\tilde S}_{z}=\sum_{i=1}^{N_q}\sigma_i^z$ up to local basis transformation, which has eigenstates satisfying 
\be{\tilde S}_{z}|J,M\rangle=M|J,M\rangle,\label{eq:eigenvalueTrappedIon}\ee 
with $M=-J,-J+1,..,0,...,J-1,J$. The maximum value of the angular momentum is $J=N_q/2$, and the state $|N_q/2,\pm N_q/2\rangle$ corresponds to all ions in the same state $|1\rangle$ or $|0\rangle$.  In this case, eigenvalue $M=-N_q/2,-N_q/2+1,...,N_q/2$. Therefore, a zero eigenvalue exists only when the total number of qubits $N_q$ is even, and the rotated subspace is spanned by the basis with the same number of ions in states $|1\rangle$ and $|0\rangle$, i.e.~the zero-eigenvalue subspace. This is slightly different from the circuit-QED implementation in the main text, where the dimension of the rotated subspace is $1$, i.e.~a subspace spanned by a single state.

As an interesting possible application of this implementation, note that in trapped-ion systems, the MS gate can be engineered to realize the Ising interaction $U_{\rm Ising}=\exp{(iJ_{ij}\sigma_x^{i}\sigma_x^{j})}$ \cite{Kim2009}. Thus we can replace $\tilde{S}_x$ in \eqref{eq:ion} with a more generalized form $\tilde S_x=\sum_i J_{0j}\sigma_i^x$, where the coupling strengths $J_{0j}$ can be tuned in situ.
In this case, our quantum walk sequence can be used to realize a quantum oracle for the NP-complete partition problem, where one seeks to partition a set of $N$ integers into two subsets with equal sum.  More explicitly, given a set $\{a_1, \ldots, a_{N_q}\}$ of integers, we set $J_{0j} = a_j$, and note that  $\tilde{S}_z=\sum_i J_{0j}\sigma_i^z = 0$ encodes a solution.  A reflection about this subspace is thus an implementation of the Grover search oracle. This approach contrasts with the proposal of Ref.~\cite{Anikeeva2021} involving central spin and central boson models with cold atoms. 

In recent years, platforms based on Rydberg atoms have also become increasingly popular for quantum information processing~\cite{Saffman2010,Morgado2021,Eldredge2017,Zeytinoglu2022}. For such systems, the Hamiltonian can be written as
\be
H_{\rm R}=\sum_{i\neq j}V_{ij}\sigma_{i}^z\sigma_{j}^z,
\ee
where $\sigma_{i}^z$ is Pauli $z$ operator for atom $i$, $V_{ij}$ is the coupling strength between atom $i$ and $j$. 
We can define
\bea
w_0&=&\exp(iH_{\rm R}t)\exp(i\phi\sigma_0^x)\exp(-iH_{\rm R}t)\nonumber\\
&=&  \exp(it\tilde{S}_z\sigma_0^z)\exp(i\phi\sigma_0^x)\exp(-it\tilde{S}_z\sigma_0^z)\label{eq:Rydbergw0}\\
&=&\exp\left[\mathrm{i}\phi\left(\cos\left(\frac{t}{2}\tilde{S}_{z}\right)\sigma_{0}^{x}-\sin\left(\frac{t}{2}\tilde{S}_{z}\right)\sigma_{0}^{y}\right)\right],\nonumber
\eea
where $\tilde S_z=\sum_i V_{0j}\sigma_i^z$.  Equation (\ref{eq:Rydbergw0}) has a similar form to \eqref{eq:ion} of trapped-ion systems.  So, following the basis transformation $\sigma_x\rightarrow\sigma_z$, $\sigma_y\rightarrow- \sigma_y$, $\sigma_z\rightarrow\sigma_x$, similar to \eqref{eq:ion-walk}, 
the walk unitary can be expressed as \be
W_0=e^{i\phi\sigma_0^x}\left(w_0e^{i\pi/2\sigma_0^x}\right),
\ee
and an interleaved sequence of $W_0$ and $S=\exp{(i2\pi/N\sigma_0^x)}$ implements the required subspace rotation.


\section{Discussions}\label{sec:discussion}
There are two potential limitations of our approach. First, the dimension of the subspace that can be rotated is limited by the connectivity of the auxiliary qubit, which is, in turn, constrained by hardware design trade-offs.  For superconducting architectures, while $N_q=4$ nearest neighbours is increasingly popular due to forward compatibility with the surface code, having connectivity significantly higher than that may not be common in the foreseeable future. However, in trapped-ion systems and Rydberg atoms, more qubits can be connected to a single ancilla.  Second, even without connectivity constraints, as the number of coupled qubits $N_q$ increases, the singular (eigen-) values are distributed in a range that increases with $N_q$, c.f. \eqref{eq:RabiFreq} and \eqref{eq:eigenvalueTrappedIon}.
It could thus be hard to make $|\cos(\Lambda_J t_g)|$ ($|\cos(\theta\lambda/2)|$) for the superconducting circuits (trapped-ion systems)
simultaneously small across the range of $\Lambda_J$, potentially necessitating taking $N$ large in order to ensure high fidelity. One approach to dealing with this is to view our approach as a special case of quantum signal processing~\cite{Low2019,Low2017}, where the many-body operator $\exp(-iHt)$ of \eqref{eq:HmatrixElement} is interleaved with single-qubit $z$ rotations of the ancilla to give 
\be
\prod_{i=1}^{2N} e^{i\phi_i \sigma_z}e^{-iHt},
\ee with vector of $z$ rotation angles $\vec{\phi} = (2\pi/N, 4\pi/N,\ldots, 4N\pi/N)$. By carefully choosing different angles $\vec{\phi}$, specific polynomial transformations of the singular values of $A$ can be engineered to give better performance over a wider range of singular values (eigenvalues) for superconducting circuits (trapped ions, Rydberg atoms). In particular, in \appref{app:signalProcessing}, we show how a vector $\vec{\phi}$ of length $10$ can be chosen which allows for subspace reflection of $6$ qubits surrounding a central ancilla, with average gate fidelity $F=0.999$ and total time less than that required by the $N=5$ topological walks for subspace rotations of four qubits.

\section{Conclusions and outlook.}\label{sec:conclusion}

We have proposed an efficient and robust method to implement rotations of a subspace. Compared with conventional decompositions of multiple-controlled-phase gates into single- and two-qubit gates, our proposal via quantum walks offers substantial speedup, enables new ways of compiling algorithms, and can serve as a building block for a wide variety of procedures. Our approach can be applied to superconducting circuits, trapped-ions systems and Rydberg atoms, without additional requirements.

Our results hinge on connections between condensed matter physics and quantum information processing, which raises the possibility that further studies on edge states and phase transitions in topological quantum walks~\cite{Kitagawa2010} may provide new perspectives on operations relevant to quantum computing.
Moreover, the interplay between quantum walks and topological phases \cite{Kitagawa2010} and non-hermitian physics \cite{Bergholtz2021} may offer new possibilities for quantum information processing.


\appendix


\section{Generalized FSBSW controlled-phase gate}\label{app:cnz}
Here, we generalize the implementation of the ${\rm{C}}_{2}\rm{Z}$ gate by Fedorov, Steffen, Baur, da Silva and Wallraff (FSBSW)~\cite{Fedorov2011} to $n$-qubits, for comparison with our quantum-walk approach.

The general idea of this implementation is to tune state $|11\rangle$ into resonance with $|02\rangle$ or $|20\rangle$. We denote this resonant Rabi frequency by $g$.  Other transitions are far off-resonance, as stated in the main text.
Since $|2\rangle$ is out of the computational space, the Rabi oscillation always starts with $|11\rangle$.
Then at time $\pi/2g$, $|11\rangle$ will be hidden in the non-computational space by the resonant Rabi oscillation, and is shielded from subsequent operations. 

\begin{figure}[hbt]
	\includegraphics[width=6cm]{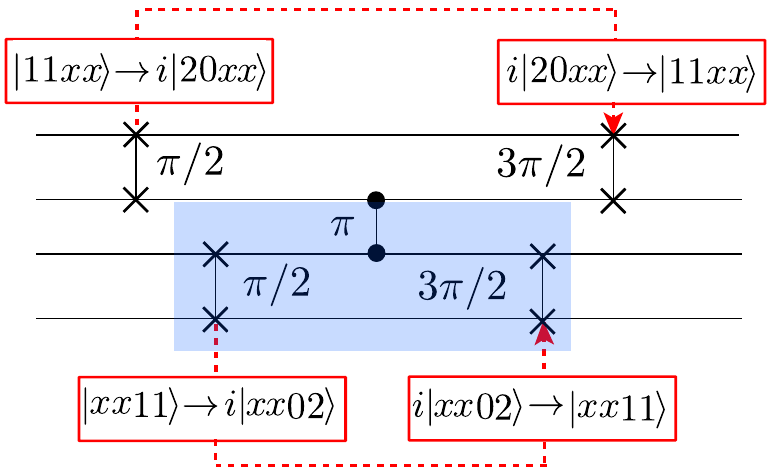}
	\caption{ Implementation of the $\rm{C}_{3}Z$ gate based on Ref. \cite{Fedorov2011}.  It effects the transformation $|0110\rangle$ to $-|0110\rangle$, with other states unaffected. Each gate is implemented via resonant oscillation between $|11\rangle$ and $|02\rangle$ ($|20\rangle$) with times $\{\pi/2,\pi/2,\pi,3 \pi/2,3 \pi/2\}$ (Rabi frequency $g$ set to 1).\label{fig:Toffoli} }
\end{figure}

First, let us focus on the area shaded in blue in \figref{fig:Toffoli}, which is the same as the gate implementation of~\cite{Fedorov2011}. In the blue area, we start with state $|xx11\rangle$, $x\in{0,1}$. This state is transferred to $|xx02\rangle$ when $gt=\pi/2$, while the remaining states are unaffected. Then a second gate 
changes $|x11x\rangle$ to $-|x11x\rangle$ when $gt=\pi$. Since $|xx11\rangle$ is first transferred to the non-computational state, we are left with $|xx00\rangle$, $|xx01\rangle$, $|xx10\rangle$.  So the second gate only imparts a $\pi$ phase shift to state $|x110\rangle$.  The third gate swaps $|xx02\rangle$ back to $|xx11\rangle$ when $gt=3\pi/2$.
In summary, this part bestows a phase shift of $\pi$ only to state $|x110\rangle$.

Now we include the gate operations on the first qubit. State $|11xx\rangle$ is shielded from subsequent operations by transferring to $|20xx\rangle$. In the blue area, we are left with states $|00xx\rangle$, $|01xx\rangle$, $|10xx\rangle$. By the previous analysis, only state $|0110\rangle$ will acquire a minus sign after the operations in the blue area. In the last step, we recover the protected state $|11xx\rangle$. So the whole gate sequence in \figref{fig:Toffoli} only takes $|0110\rangle$ to $-|0110\rangle$. 

If we apply NOT gates to qubits 2 and 3 in the  initial and final stages, the whole sequence shown in \figref{fig:Toffoli} bestows a phase factor -1 only to state $|0000\rangle$, which is the same gate operation implemented in the main text.

To add an additional qubit, the trick is to swap state $|11xxx\rangle$ to the non-computational state $|20xxx\rangle$ by coupling to the additional qubit. Then the subsequent gate operation only changes $|x1110\rangle$ to $-|x1110\rangle$.  This gate operation can be realized by applying NOT gates to the first qubit before and after the gate sequence shown in \figref{fig:Toffoli}. Finally, swapping state $|11xxx\rangle$ back completes the operation. This five-qubit gate then only changes state $|01110\rangle$ to $-|01110\rangle$.  Repeating the above procedure, we can implement a ${\rm{C}}_{n-1}\rm{Z}$ gate at a total CZ time cost of $(2n-3)\pi/g$.  Similar results are also obtained in Ref.\cite{Nikolaeva2021}. 



\section{Subspace reflection via quantum signal processing}\label{app:signalProcessing}

Here we show how quantum signal processing can be used to reflect the subspace spanned by the left eigenstates of $A$. This method can be efficiently applied to a larger number ($6$ vs $4$) of qubits than the $N=5$ method based on topological walks presented from the main text, and requires less total time to implement with no loss in fidelity. We consider an ideal implementation in superconducting systems, with topological walk sequence $W_{k}^{[2N,1]}$ of \eqref{eq:sequenceCircuit}, taking $k = 0$. With $N=5$, it can be seen (see next section) that 
$$\bra{i} W_{k=0}^{[10,1]}(t)\ket{i} = \bigoplus_{J}\bpm P_{tw}\lp \cos(\Lambda_J t)\rp\epm,$$ 
where $P_{tw}(x)$ is the degree $10$ polynomial given by $P_{tw}(x) = 2x^{10} - 1$.  This polynomial has the desired property that $P_{tw}(\cos(\Lambda t) = 1$ when $\Lambda = 0$, and $P_{tw}(\cos(\Lambda t)) \approx -1$ when $\cos(\Lambda t)$ is small. 

\begin{figure}[hbt]
	\includegraphics[width=9cm]{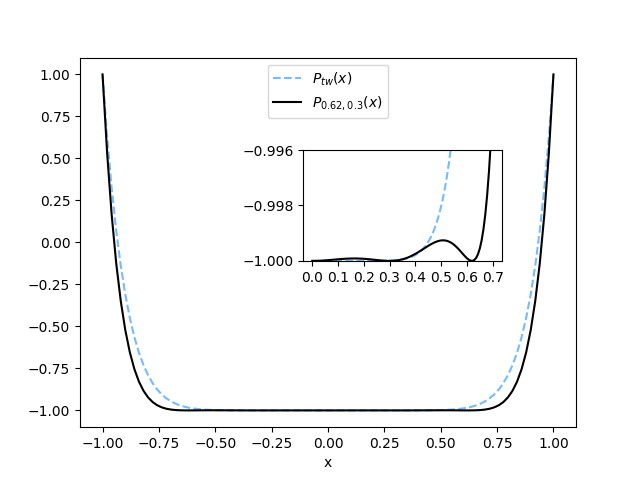}
	\caption{Polynomials $P_{tw}(x)$ and $P_{0.62,0.3}(x)$ corresponding to the $N=5$ topological walk and the quantum signal processing procedure described in the text, respectively. Inset: close-up of the behaviour of the polynomials in the range $x\in[0,0.7]$.}\label{fig:polynomials} 
\end{figure}

Consider instead, a different degree $d=10$ polynomial 
\be
P_{a,b}(x) = 2x^2\frac{(x^2-a^2)^2(x^2-b^2)^2}{(1-a^2)^2(1-b^2)^2}-1,\label{eq:poly-transform}
\ee
with $a=0.62, b=0.3$  This takes the value $1$ at $x=\pm 1$ and rapidly decays to near $-1$ away from these two extremes (see Fig.~\ref{fig:polynomials}). Furthermore, $P_{0.62,0.3}(x)$ satisfies 
\begin{enumerate}
	\item $\forall x\in[-1,1] : \abs{P_{0.62,0.3}}\le 1$
	\item $\forall x\in(-\infty,-1] \cup [1,\infty) : \abs{P_{0.62,0.3}}\ge 1$
	\item $\forall x\in\mathbb{R} : P_{0.62,0.3}(ix) P_{0.62,0.3}^*(ix)\ge 1$
\end{enumerate}
and thus, by Theorem 4 of~\cite{Gilyen2019} there exists $\vec{\phi}=(\phi_0, \phi_1, \ldots, \phi_d)$ such that 
\be
\bra{i}\left[\prod_{j=1}^{d} \left( e^{i\phi_{d+1-j} \sigma_z}W(x)\right)\right] e^{i\phi_0\sigma_z} \ket{i}= P_{0.62,0.3}(x )
\ee
where $i\in\{0,1\}$ and $W(x) = \begin{pmatrix} x & -i\sqrt{1-x^2}\\ -i\sqrt{1-x^2} & x \end{pmatrix}$. The angles $\vec{\phi}$ can be found efficiently, e.g.~by a method presented in~\cite{Gilyen2019}, which we compute to be $\vec{\phi} = (\eta_1, \eta_2, -\eta_2, \eta_1, 0, \eta_1, \eta_2, -\eta_2, \eta_1, 0,0)$,
where $e^{i\eta_1} = 0.8718 + 0.4899i$, $e^{i\eta_2} = 0.3831 +0.9237i$.

Taking $x=\cos(\Lambda_J t)$ and comparing to \eqref{eq:approxR} 
from the main text shows that, interleaving $e^{-iHt}$ with single qubit $z$ rotations with angles given by $\vec{\phi}$ effects the desired reflection of the (left) singular vectors of $A$. Now consider up to $6$ qubits each coupled to a central ancilla.  With homogenous couplings $g_i = g$ the singular values of $A$ satisfy  $\Lambda_J/g \in \{1,\sqrt{2},\sqrt{3}, 2, \sqrt{5}, \sqrt{6}\}$.  Taking $t = 0.88/g$ is sufficient to give an average gate fidelity $F = 0.999$.  Compare this to the topological walk method with  $N=5$, which requires the same number (ten) of applications of $e^{-iHt}$ to achieve $F=0.999$ when four neighbouring qubits are coupled to the central ancilla, but each application of $e^{-iHt}$ is applied for time $\pi/(3g)$.

A similar approach can be taken with ion trap or Rydberg implementations (see \secref{sec:trappedIon}).  For example, consider $6$ ions collectively coupled to an ancilla, which has quantum walk unitary given by~\eqref{eq:ion-walk}, where $\lambda\in\{-3,-2,-1,0,1,2,3\}$.  Taking the MS gate parameter $\theta = \pi/2$ gives $\cos(\theta \lambda / 2) \in \{-1/\sqrt{2}, 0, 1/\sqrt{2}, 1\}$/ Define the degree $6$ polynomial
\be
P_{a}(x) = 2x^2\frac{(x^2-a^2)^2}{(1-a^2)^2}-1,\label{eq:poly-transform}
\ee
with $a=1/\sqrt{2}$, which satisfies $P_a(1) = 1$, and $P_a(\pm 1/\sqrt{2}) = -1$. Thus, $6$ applications of the walk operator~\eqref{eq:ion-walk} interleaved with single qubit rotations is sufficient, in the ideal case, to implement the desired reflection with fidelity $1$.

\subsection{Polynomial transformations induced by topological walks}\label{app:poly-transform}

Let $N$ be an odd integer. Define $W(x) = \begin{pmatrix} x & -i\sqrt{1-x^2}\\ -i\sqrt{1-x^2} & x \end{pmatrix}$ and $R_j = \begin{pmatrix} \omega_j & 0\\ 0 & \omega_j^* \end{pmatrix}$, where  $\omega_j = e^{2\pi i j/N}$ for $j \in \{1,\ldots, N\}$.  Define
\begin{align}
	W_N(x)  &= \prod_{j=1}^NR_{N+1 - j}W(x) \\
	&= W(x) \prod_{j=2}^NR_{N+1 - j}W(x)
\end{align}
where the second line follows from $R_N = \mathbb{I}$. 
Here we show that $\bra{i} \left[W_N(x)\right]^2\ket{i} = 2x^{2N} - 1$.

From~\cite{Gilyen2019}, $W_N(x)$ takes the form
\be
W_N(x) = \begin{pmatrix} P(x) & iQ(x)\sqrt{1-x^2} \\ iQ^*(x)\sqrt{1-x^2} & P^*(x)\end{pmatrix}
\ee
where $P,Q\in \mathbb{C}[x]$ are polynomials of maximum degree $N$ and $N-1$ respectively that satisfy $\abs{P(x)}^2 + (1-x)^2\abs{Q(x)}^2 = 1$ for all $x\in[1,1]$. Furthermore, $P$ and $Q$ have parity $N \bmod 2$ and $(N-1) \bmod 2$, respectively. 

We now show that $P(x) = \bra{0} W_N(x) \ket{0} = x^N$, from which it follows that $\bra{i}\left[W_N(x)\right]^2\ket{i} = 2x^{2N} - 1$.  First note that $P(x)$ is a polynomial in $x$ and $(-i\sqrt{1-x^2})^2$ (see the proof of Lemma 2 of the Supplementary Material of~\cite{Cedzich2013}). For real $x$, it is therefore invariant to the transformation $W(x) \mapsto W^*(x)$. Then, observing that $R_j = R^*_{N+2-j}$, we have
\begin{align}
	\bra{0} W_N(x) \ket{0}^*  &= \bra{0}W^*(x)\prod_{j=2}^NR^*_{N+1 - j}W^*(x)\ket{0} \\
	&=  \bra{0}W^*(x)\prod_{j=1}^{N-1}R^jW^*(x)\ket{0} \\
	&= \bra{0}W(x)\prod_{j=1}^{N-1}R^jW(x)\ket{0}  \\
	&= \bra{0}\left[W_N(x)\right]^\top\ket{0} \\
	&= \bra{0}W_N(x)\ket{0}
\end{align}
and thus, $P(x)$ is real. From Lemma 2 of the Supplementary Material of~\cite{Cedzich2013}, for $N$ odd, the trace of $W_N$ satisfies
\be
\tr {W_N(x)} = 2x^N
\ee
As $P(x)$ is real, it follows that $P(x) = x^N$.

\section{Fidelity for general rotation angles}\label{app:rotation-fidelity}

In \secref{sec:benchmark} we computed the $N=3,5,7$ average gate fidelities, for the case where $k=0$, i.e. subspace reflections.  Here we investigate the average gate fidelities for general rotation parameters $k$. For concreteness we consider a central ancilla connected to $N_q=4$ neighbour qubits, and a walk of length $2N$, where $N=3$. Explicit calculation gives 
\begin{align}
M(k,t) &:= \bra{1_0}W^{[6,1]}_{ k}(t)\ket{1_0}\\
&= \bigoplus_{J\in\{0,1\}^4} \bpm f(k, \Lambda_J t) \epm
\end{align}
where
\begin{widetext}
\begin{align}
f(k,\lambda):=&e^{6ik}\cos^6(\lambda)+\sin^2(\lambda)\cos^4(\lambda) \left[ (e^{4ik }- e^{-4ik})- (e^{2ik}-e^{-2ik}) - 3\right] \\
&- 3\sin^4(\lambda)\cos^2(\lambda)-\sin^6(\lambda)
\end{align}
\end{widetext}
and where $\Lambda_J$ are given by~\eqref{eq:RabiFreq}, and thus $\Lambda_J/g$ take values in the range $\Lambda_J/g \in \{1,\sqrt{2},\sqrt{3}, 2, \sqrt{5}, \sqrt{6}\}$. Taking $U_{\rm ideal}=\mathsf{diag}(e^{6ik},-1,-1,\ldots -1)$ gives an average gate fidelity of
\be
F=\frac{\abssq{\tr{M U_{\rm ideal}^\dag}} + \tr{M^\dag M}}{n(n+1)}, \ee
where $n=2^{N_q}$ and where
\begin{align} 
\tr{M^\dag M} &= 1 + \sum_{J:\Lambda_J\neq 0} \abs{f(k,\Lambda_J t)}^2,\\
\abs{ \tr{U_{\rm ideal}^\dag M}}^2 &= \abs{1-\sum_{J:\lambda_J\neq 0}f(k,\Lambda_J t)}^2.
\end{align}

\begin{figure}[hbt]
	\includegraphics[width=9cm]{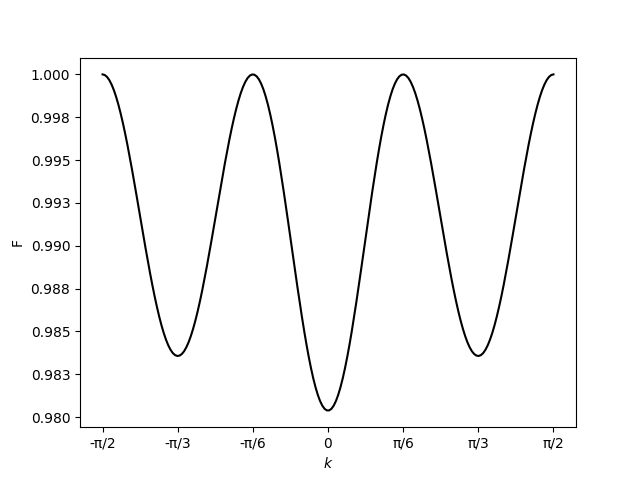}
	\caption{Average gate fidelity for subspace rotation via topological walk, as a function of rotation parameter $k$. Values correspond to $N_q=4$, $N=3$ and $t=0.333\pi/g$. }\label{fig:rotation-fidelity} 
\end{figure}

These fidelities are show in Fig.~\ref{fig:rotation-fidelity}, taking $t = 0.333\pi /g$ as is the main text. The average gate fidelity varies slightly with $k$, with maxima and minima corresponding to values of $k$ leading to constructive or destructive interference in the expression for $f(k,\lambda)$ above.

\bibliography{library.bib}

\end{document}